\documentclass[traditabstract]{aa} 
\usepackage{graphicx}
\usepackage{natbib}
\usepackage{deluxetable}
\usepackage[pdftex,                
bookmarks=true,                    
bookmarksnumbered=true,            
colorlinks=true,            
citecolor=blue,         
linkcolor=blue,     
menucolor=blue,         
urlcolor=cyan,       
linkbordercolor={0 0 1},           
pdfborder= {0 0 1},
frenchlinks=True]{hyperref}
\usepackage{txfonts}
\providecommand{\eprint}[1]{\href{http://arxiv.org/abs/#1}{#1}}
\providecommand{\adsurl}[1]{\href{#1}{ADS}}
\newcommand{\fig}[5]{
        \begin{figure}[!bt]
        \begin{centering}
        \includegraphics[#3]{#2}
      \end{centering}
      \renewcommand{\baselinestretch}{1}
        \vspace*{-.3in}
        \caption[#4]{#5}
        \label{fig:#1}
        \end{figure}}

\def\deg{\ensuremath{^{\circ}}}
\def\micron{\ensuremath{\mu m}}

\def\tbright{\ensuremath{1950^{+260}_{-190}\textrm{~K}}}  
\def\pflux{\ensuremath{0.44^{+0.12}_{-0.08}\textrm{~mJy}}}

\begin{document}

   \title{ACME Stellar Spectra}

   \subtitle{I. Absolutely Calibrated, Mostly Empirical Flux Densities
     of 55~Cancri and its Transiting Planet
     55~Cancri~e}

   \author{Ian J. M. Crossfield
          \inst{1}\fnmsep\inst{2}
          }

          \institute{
             Max-Planck Instit\"ut f\"ur Astronomie, K\"onigstuhl 17, 69117, Heidelberg, Germany \\
         \and
            Department of Physics \& Astronomy, University of
            California Los Angeles, Los Angeles, CA 90095, USA \\
            \email{ianc@astro.ucla.edu}
             }

   \date{Sent to publisher: \today}

 
  \abstract
  {The ACME Spectra project provides absolutely calibrated, mostly
    empirical spectra of exoplanet host stars for use in analysis of
    the stars and their planets.  Spectra are obtained from
    ground-based telescopes and are tied directly to calibrated
    ground- and space-based photometry.  The spectra remain only
    ``mostly'' empirical because of telluric absorption, but
    interpolation of stellar models over the gaps in wavelength
    coverage provides continuous stellar spectra. Among other uses,
    the spectra are suitable for precisely converting observed
    secondary eclipses (occultations) into absolute flux units with
    minimal recourse to models.  In this letter I introduce ACME's
    methods and present a calibrated spectrum of the nearby,
    super-Earth hosting star 55~Cancri that spans the range from
    0.81--5.05\,\micron.  This spectrum is well-suited for
    interpreting near- and thermal-infrared eclipse observations.
    With this spectrum I show that the brightness temperature of the
    small, low-mass transiting planet 55~Cnc~e is \tbright\ at
    4.5\,\micron\ (cooler than previously reported), which corresponds
    to a planetary flux of \pflux. This result suggests the planet has
    some combination of a nonzero albedo, a moderately efficient
    redistribution of absorbed stellar irradiation, and/or an
    optically thick atmosphere, but more precise eclipse measurements
    are required to distinguish between these scenarii.}
   { }
   { }
   { }
   { }

   \keywords{infrared: stars --- planetary systems ---
     stars:~individual (\object{55~Cnc}) --- stars: fundamental
     parameters --- techniques: spectroscopic --- planets and
     satellites:~individual (\object{55~Cnc~e})}

   \maketitle
%

\section{Introduction}

Measurements of flux densities are the primary diagnostic of
conditions in stellar and planetary atmospheres, and are used to
constrain albedos, sizes, temperatures abundances, temperature
profiles, and even interior conditions of stars and planets. 
%
In the last decade observations of transiting planets during eclipse
have begun to reveal the intrinsic emission of these externally
irradiated bodies \citep[e.g.,][]{deming:2005,charbonneau:2005}.
Though initially confined to space-based observatories such as Spitzer
\citep[and, more recently, Hubble;][]{swain:2012}, more favorable
targets and improved observing strategies have recently enabled a
growing number of ground-based measurements
\citep[e.g.,][]{sing:2009ogle,demooij:2009}.

Because eclipse measurements are inherently relative, precise
planetary fluxes require stellar fluxes yet more precise. The state of
the art is now reaching the point at which this condition is no longer
met; for example, recent measurements of eclipses with Spitzer/IRAC
have attained relative precisions of $1-4\%$
\citep[e.g.,][]{agol:2010, campo:2011}, comparable to the instrument's
3\% absolute calibration accuracy \citep{SSC:2012}.

The goal of the ACME project is to provide a catalog of Absolutely
Calibrated, Mostly Empirical (ACME) spectra of exoplanet host
stars with high accuracy and precision.  In the rapidly approaching
era of exoplanet spectroscopy \citep[e.g.,][]{swain:2012} it is
imperative that such an effort advance beyond standard photometric
calibrations and focus on calibrated spectroscopy. In addition to the
aforementioned conversion of planetary eclipse depths into absolute
flux units, other potential uses for such a catalog include the study
of the planets' host stars: abundance analyses, spectral typing, and
refined bolometric luminosities and effective temperatures are all
enabled by such data.

I describe the observation, reduction, and calibration of ACME spectra
in \S~\ref{sec:obs}. As a particular case study I present a calibrated
spectrum of the nearby star \object{55~Cancri} (\object{55~Cnc}) in
\S~\ref{sec:results}\footnote{Spectral data are available in electronic
       form at the CDS via anonymous ftp to cdsarc.u-strasbg.fr
       (130.79.128.5) or via
       \url{http://cdsweb.u-strasbg.fr/cgi-bin/qcat?J/A+A/} }.  With this calibrated spectrum I provide an
improved measurement of \object{55~Cnc~e}'s flux at 4.5\,\micron\ in
light of recent observations \citep{demory:2012}, and I conclude with
future prospects.


\section{Observations and Methods}
\label{sec:obs}
%

\subsection{Summary of Observations}
Observations thus far have been made solely with SpeX
\citep{rayner:2003}, a cryogenic, cross-dispersed, near-infrared
echelle spectrograph at the NASA InfraRed Telescope Facility
(IRTF). Observations and reduction of SpeX data generally follow the
steps used in compilation of the IRTF Spectral Library
\citep[][]{cushing:2005,rayner:2009}; below I review briefly these
steps and highlight the differences between my analysis and that of
the IRTF Library's.

All objects are observed using SpeX's short-wavelength (SXD)
mode. This mode offers continuous wavelength coverage from
0.8--2.4\,\micron\ (except for the gap at 1.81--1.87\,\micron) at a
spectral resolution of 2,000 for the standard 0.3'' slit.  For bright
targets such as 55~Cnc ($K_s=4$), wavelength coverage can be extended
into the thermal infrared using one of the several long-wavelength
(LXD) modes, which offer a resolution of 2,500.

For each target, observations of a nearby A0~V star allows telluric
calibration \citep{vacca:2003}; these stars are typically chosen to be
located within $15\deg$ and 0.1~airmass unit of the primary
observation. Observations are taken at the parallactic angle to avoid
spurious spectral slopes induced by differential atmospheric
refraction.
I use the standard SpeX calibration macros to generate the dark, flat,
and arc frames used to calibrate the data.
Table~\ref{tab:log} shows a log of the observations of 55~Cnc
discussed in \S~\ref{sec:results}.

\subsection{Initial Spectral Reduction}
I use the excellent SpeXTool software package \citep{cushing:2004} to
reduce, combine, and clean the data, a process that consists of the
following steps\footnote{The SpeXTool procedures used, in order, are:
  \texttt{XSPEXTOOL}, \texttt{XCOMBSPEC}, \texttt{XTELLCOR},
  \texttt{XLIGHTLOSS} (when appropriate), and \texttt{XMERGEORDERS}.}.
First, optimal extraction converts the differences of pairs of nodded
frames into uncalibrated spectra in flux density units. I use a
constant background level and large apertures (2-3'' radii) in the
extraction; note that the particular choice of reduction parameters is
retained in the headers of the FITS files generated by SpeXTool.
Individual spectra for an object are then combined using a robust
weighted mean, and the mean target spectrum is corrected for telluric
absorption using the mean A0~V spectrum \citep{vacca:2003}.  If an
object was not observed at the parallactic angle, I correct the
spectrum for losses resulting from chromatic seeing and atmospheric
dispersion.
Finally, the individual echelle orders are combined by applying small
($\lesssim 2\%$) offsets to the individual orders.  The final result
is a spectrum with well-determined uncertainties at all
wavelengths. In most cases an additional step is the trimming of
pixels located in regions of extremely low telluric transmission or
with $\textrm{S/N}<5$.

Note that unlike objects catalogued in the IRTF Spectral Library
\citep{cushing:2005,rayner:2009} the spectra of ACME targets are not
shifted to zero radial velocity; but such corrections can easily be
made by using the copious body of high-precision radial velocity
characterization of these systems
\citep[e.g.,][]{fischer:2008}. Future ACME spectra will be corrected
for interstellar reddening, but owing to its proximity 55~Cnc requires
no such correction \citep{vonbraun:2011}.

\subsection{Spectral Calibration}
To place an extracted spectrum on an absolute flux scale I follow
\cite{rayner:2009}, whose Eq.~1 computes the correction factor $C_X$
for each bandpass $X$ by comparing to catalog photometry.
These factors provide the ratios by which the uncalibrated spectrum
should be scaled in each bandpass: multiplying the spectrum by the
weighted average of the $C_X$ calibrates the flux scale of the
spectrum.  Fainter objects observed at $<2.5\,\micron$ use only 2MASS
photometry \citep{skrutskie:2006} and the spectral response curves
from \cite{cohen:2003}; targets observed at longer wavelengths (such
as 55~Cnc) also use WISE W1 and/or W2 photometry \citep{wright:2010}
and the spectral response curves from \cite{jarrett:2011}.

Telluric absorption causes several gaps in the observed spectrum: in
addition to several smaller regions, the largest gaps span
1.81--1.87\,\micron, and (for longer-wavelength data) from
2.50--2.95\,\micron\ and 4.18--4.59\,\micron.  These last two gaps
intersect the WISE W1 and W2 response curves and so must be accounted
for before flux calibration can take place; this step is also required
to best constrain bolometric luminosities and for future application
to space-based observations not necessarily restricted to telluric
windows. As described below, to fill these gaps I use the BT-Settl
library\footnote{Available online at
  \url{http://phoenix.ens-lyon.fr/}} \citep{allard:2010}, which
provides high-resolution model spectra across a wide range of
$T_\textrm{eff}$, $\log g$ and [Fe/H].

I interpolate the model spectra to obtain a spectrum with the stellar
parameters appropriate for the target star; for 55~Cnc, these are
$5196\pm24$~K, $\log g \textrm{(cgs)} = 4.45\pm0.01$, and
$\textrm{[Fe/H]}=0.315\pm0.03$ \citep{valenti:2005,vonbraun:2011}.
Convolution of the resultant model with a Gaussian profile brings the
model to an approximate resolution of 2,000, and computation of $C_X$
factors for the model place it on the same flux scale as the observed
spectrum.  At this point the observed and model spectra sometimes
exhibit mismatched spectral slopes at gap edges. To mitigate these
discrepancies I measure the weighted mean flux density in the 10~nm
adjacent to the gap boundary for both model and observed spectra; from
these measurements I compute a weighted linear fit that I apply to the
model spectrum.  Thus the details of spectral features in the model
are retained, while the spectral slope is adjusted to match the
observations.  I re-compute the weighted mean of the $C_X$ factors
using the continuous, hybrid spectrum; the uncertainty on this
quantity provides an estimate of the uncertainty in the absolute flux
calibration.  Applying this mean factor to the data produces an
absolutely calibrated, mostly empirical spectrum of the target star.

The situation is slightly more complicated for 55~Cnc, which exceeds
the nominal WISE saturation limits in the W1 and W2 bands.  For a star
of this magnitude, WISE overestimates the flux in the W1 and W2 bands
by, respectively, 0.1~mag and 0.8~mag when compared to fainter stars
\citep[Sec.~VI.4 of ][updated 2011 April~26]{cutri:2012}. I therefore
apply these offsets to the reported magnitudes before flux-calibrating
the spectrum.  

As an additional check, I measure 55~Cnc's flux at 3.6\,\micron\ and
4.5\,\micron\ in archival IRAC photometry (Spitzer Programs 48, 33970,
40976, 70076, and 80231; P.I.s G.~Fazio, K.~Luhman, M.~Marengo, K.~Su,
and M.~Gillon, respectively).  I determine the stellar flux densities
in these bands to be $7.2\pm0.3$~Jy and $4.3\pm0.1$~Jy, where the
uncertainties represent the spread in values across several observing
programs.  These flux values exceed the WISE catalogue fluxes (after
applying corrections for the slightly different isophotal wavelengths
and for partial saturation, as described above) by 22\%.  


Both the IRAC and WISE measurement pairs are internally
self-consistent, as demonstrated by the fact that both sets yield
$\frac{F_{\nu,1}}{F_{\nu,2}} \left( \frac{\lambda_1}{\lambda_2}
\right)^{2}$ equal to 1.05, near the value of unity expected near the
Rayleigh-Jeans limit. The deciding factor is the comparison with the
calibrated model spectrum of \cite{vonbraun:2011} described in the
following section.  Spectral calibration using the WISE photometry
provides an excellent agreement with \citeauthor{vonbraun:2011}'s
model spectrum, but the agreement becomes worse by a factor of $\sim$8
when calibrating with the IRAC photometry.

The 2MASS $K_s$ band measurement of 55~Cnc has a quality flag of $E$,
indicating a poorly determined flux; therefore I do not use this
bandpass for flux calibration. In addition, the final calibration
suggests that the 2MASS $J$ and $H$ measurements underestimate the
stellar flux by roughly 25\%. The calibration is dominated by the
smaller uncertainties of the WISE photometry, and I use $J$, $H$, W1,
and W2 in the final calibration.  55~Cnc's calibrated spectrum is
shown in Fig.~\ref{fig:sed} and is available in electronic
form. Comparison of this spectrum with the photometrically-calibrated
\cite{pickles:1998} spectrum of \cite{vonbraun:2011} indicates a good
match between the two, as described below.


\section{Results: Stellar and Planetary Fluxes}
\label{sec:results}

\subsection{Spectrum and Bolometric Flux of 55~Cnc}
The star 55~Cnc hosts five planets \citep{fischer:2008,dawson:2010},
including the small transiting planet 55~Cnc~e
\citep{winn:2011,demory:2011,gillon:2012}.  Because of these planets,
and by virtue of its proximity, the star has been the subject of
detailed characterization.  Recently \cite{vonbraun:2011} used new
interferometric and archival photometric measurements to determine the
star's bolometric flux \citep[using models from][]{pickles:1998} to
be $(1.227 \pm 0.0177) \times 10^{-10} \textrm{~W~m}^{-2}$. The
calibrated spectrum of 55~Cnc shown in Fig.~\ref{fig:sed} emits a flux
consistent with that of \citeauthor{vonbraun:2011}'s spectrum from
0.81--2.49\,\micron: the difference between the two is only $(1.8 \pm
7.2) \times 10^{-12}\textrm{~W~m}^{-2}$, which independently confirms
the calibration of both spectra.  

In the IRAC1 and~2 bands the flux densitites of 55~Cnc are
$6.153\pm0.006$~Jy and $3.444 \pm 0.006$~Jy, respectively, where these
uncertainties represent only the precision (not the accuracy) of the
measurements.  The formal uncertainty on these values is in fact
dominated by the absolute calibration of the ACME spectrum: the
weighted mean of the $C_X$ factors for 55~Cnc has an uncertainty of
12\%.  Considering this level of accuracy, the stellar fluxes are
  in good agreement with the stellar flux densities inferred solely
  from the calibrated BT-Settl model: 6.03~Jy and 3.78~Jy,
  respectively.  This situation will improve for fainter stars (which
avoid saturation and are measured more precisely by 2MASS and WISE),
but even here the close correspondence (1.4\%) between my and
\citeauthor{vonbraun:2011}'s in-band fluxes suggests that the true
accuracy of 55~Cnc's spectrum may be better than 12\%.


\subsection{Brightness Temperature and Flux of 55~Cnc~e}
 By dint of its high temperature and its host star's apparent
  brightness, and despite its small transit depth, 55~Cnc~e is the
  planet whose intrinsic emission spectrum is most amenable to
  characterization \citep{winn:2011,gillon:2012}. At least in the near
  term, this planet will be our sole source of detailed information as
  to the nature of small, highly irradiated planets and their
  atmospheres.

The mass, size, and density of 55~Cnc~e are all intermediate between
the Solar System's terrestrial and gas giant planets
\citep{dawson:2010,gillon:2012}, so its composition cannot be
determined solely on the basis of these bulk properties
\citep[cf. ][]{valencia:2010}. However, models of planetary interiors
suggest that 55~Cnc~e hosts a substantial envelope of volatile
materials \citep{gillon:2012}. The planet's mass and radius can be fit
with an Earthlike core and an envelope of either a H$_2$/He
($\sim0.03\%$ by mass) or H$_2$O ($\sim20\%$)
\citep{valencia:2010}. Models of atmospheric mass loss for CoRoT-7b
\citep[a smaller, denser, and hotter planet;][]{leger:2009} reveal
that it is difficult for such a small, highly irradiated planet to
retain substantial amounts of H$_2$/He
\citep{valencia:2010,heng:2012},  and models of sublimated rock
  atmospheres suggest that even heavier species will not remain aloft
  beyond the highly irradiated dayside \citep{castan:2011}. Thus it
is unlikely that 55~Cnc~e's volatile component is a H$_2$/He envelope,
and it seems more likely to be made up of heavier species.

At present, only observations during transit and/or eclipse can
constrain the atmospheric properties of such a planet
\citep[e.g.,][]{miller-ricci:2009}.  The first thermal emission from
55~Cnc~e has recently been reported \citep{demory:2012} via detection
with Spitzer/IRAC at 4.5\,\micron\ of an eclipse whose depth is only
$131\pm28$~ppm \citep{demory:2012}.  This result provides a timely
opportunity to demonstrate the utility of ACME spectra: I now
demonstrate that the planet's brightness temperature, $T_B$, is cooler
than the initially reported value of $2360\pm300$~K
\citep{demory:2012}.

The brightness temperature $T_B$ of an eclipsing planet can be
determined by solving the following relation:
\begin{equation}
\frac{F_p}{F_*} =  \frac{\int B_\lambda (T_B) S_{X} (\lambda) d \lambda}{\int F_{\lambda}^{*}  S_X (\lambda) d \lambda} 
\left( \frac{R_p}{R_*} \right)^2
\end{equation}
Here $F_\lambda^*$ is the stellar flux density, $S_{X}(\lambda)$ is
the system response in bandpass $X$, $B_\lambda$ is the Planck
function, and the other symbols have their usual meanings. In a
typical eclipse observation the planet/star radius ratio is known to
high precision from transits and the planet/star flux ratio is
measured from eclipses.  The ACME Spectra project provides
$F_\lambda^*$, so $T_B$ is the only unknown quantity.

Combining the eclipse depth \citep{demory:2012}, planetary properties
from \cite{gillon:2012}, stellar properties from \cite{valenti:2005}
and \cite{vonbraun:2011}, the stellar spectrum shown in
Fig.~\ref{fig:sed}, and propagating the uncertainties in all these
quantities using a Monte Carlo approach yields $T_B =$\tbright\
(corresponding to \pflux), lower than previously reported  by
  400~K.  If 55~Cnc has an exposed, rocky component then this new,
  lower temperature suggests that atomic Fe would be more prevalant in
  the planet's Na-dominated atmosphere than SiO, and in general
  outgassed materials would be rather less common 
  than in the hotter case \citep{castan:2011,miguel:2011}. 

Using a blackbody for $F_\lambda^*$ instead of the stellar
spectrum I obtain the published value of $T_B = 2360^{+350}_{-240}$, a
result which emphasizes the point that a star does not emit the same
flux density as a blackbody at the stellar effective temperature.
However, note that in converting $T_B$ into a flux density I have
assumed the planet emits like a blackbody in the IRAC2 band; this is
probably not the case, but it is forgivable considering our current
ignorance of the planet's true emission spectrum.

This new and improved temperature estimate is mildly lower (by
$1.4\sigma$) than the 2300~K expected in the zero albedo, zero
redistribution case discussed by \cite{demory:2012}.  Kepler-10b is
another small, low-mass, highly irradiated planet whose albedo has
been inferred from its optical phase curve to be $0.48 \pm 0.35$
\citep{batalha:2011,rouan:2011}. It is intriguing to note that such an
albedo for 55~Cnc~e would make the 4.5\,\micron\ brightness
temperature quite consistent with the no-redistribution case
(1930~K). However, this agreement is probably coincidental: in
contrast to 55~Cnc~e, Kepler-10b contains few or no volatiles
\citep{batalha:2011} and is presumably of an altogether different
nature.

The large uncertainty on the 4.5\,\micron\ eclipse depth precludes any
unambiguous inferences as to the nature of 55~Cnc~e's emission
spectrum.  However, if the brightness temperature reported here is
characteristic of the temperature in the planet's atmosphere, some
combination of the following three scenarios may therefore be
required. The planet has a nonzero albedo, caused either by clouds
\citep[][but see also
\citeauthor{leger:2011}~\citeyear{leger:2011}]{schaefer:2009,
  castan:2011} or perhaps by the planet's surface; the planet exhibits
some redistribution of absorbed stellar irradiation, suggesting a
dense atmosphere \citep[a more tenuous atmosphere or a lava ocean
cannot transport the requisite heat flux;][]{castan:2011, leger:2011};
or the planet has an optically thick atmosphere, of which the IRAC2
observations probe a particular layer (one not necessarily indicative
of the planet's surface or equilibrium temperature).  Further
observations at higher precision are of the utmost important if we are
to distinguish between these scenarii.

These results demonstrate the utility of calibrated stellar spectra
for determining the properties of transiting planets and their host
stars.  In particular, the spectrum presented here should be of
  great utility in planning and interpreting future observations of
  55~Cnc~e with Spitzer/IRAC and the Hubble Space Telescope.  A
forthcoming paper will present the first large batch of approximately
one dozen ACME spectra, several of which extend into the thermal
infrared.  Secondary eclipses have been observed for all these
targets, and such calibrated spectra should provide a useful tools
with which to study these systems.

\begin{acknowledgements} 

  I thank T.~Guillot and the anonymous referee for insightful comments
  and suggestions which improved the quality of this paper.  I was
  supported by the UCLA Dissertation Year Fellowship and by EACM.
  This research has made use of the Exoplanet Orbit Database at
  \url{http://www.exoplanets.org}, the Extrasolar Planet Encyclopedia
  Explorer at \url{http://www.exoplanet.eu}, and free and open-source
  software provided by the Python, SciPy, and Matplotlib communities.
  I will gladly distribute my raw data products, and many of my
  algorithms, to interested parties upon request.
\end{acknowledgements}


\clearpage

\begin{deluxetable}{ c c c c c c c c}
\tablecaption{SpeX Observations of 55~Cnc\label{tab:log}}
\tablewidth{0pt}
\tablehead{
 \colhead{Spectral} & \colhead{UT Date} & \colhead{Spectroscopy}  & \colhead{R}  & \colhead{Exp. Time}  & \colhead{A0 V}  & \colhead{Airmass} & \colhead{Seeing}\\
 \colhead{Type} & \colhead{} & \colhead{Mode} & \colhead{} & \colhead{(sec)} & \colhead{Standard} & \colhead{Range}  & \colhead{(arcsec)}
}
\startdata
G9 IV & 2012 Apr 22 & SXD     & 2000 & 400  & HD 71906 & 1.22-1.25 & 1.0 \\ 
     & 2012 Apr 22 & LXD2.3  & 2500 & 66   & HD 71906 & 1.14-1.16 & 0.75 \\ 
\enddata

\end{deluxetable}

\fig{sed}{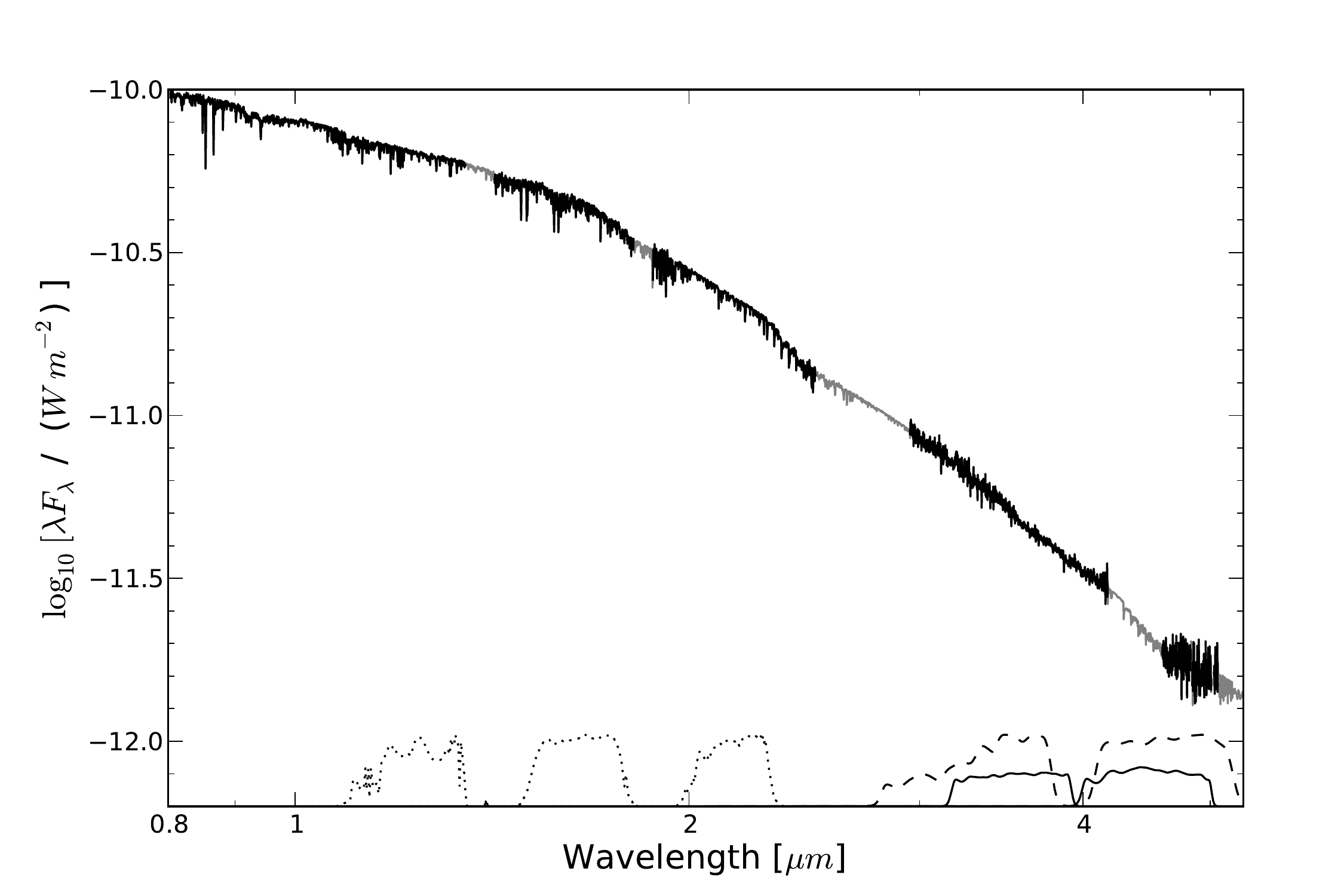}{width=6in}{}{Absolutely calibrated, mostly
  empirical spectrum of 55~Cnc. The data in black are measured, while
  those in gray are modeled.  At bottom are the response curves from
  2MASS \citep[dotted;][]{cohen:2003}, WISE
  \citep[dashed;][]{jarrett:2011}, and Spitzer/IRAC \citep[solid;
  ][]{SSC:2012}.}

\end{document}